\documentclass[
  aps, reprint, superscriptaddress,
  amsmath,amssymb
]{revtex4-2}

\usepackage[english]{babel}
\usepackage{bbold}
\usepackage{bbm}
\usepackage{color}
\usepackage{rotating}
\usepackage{graphicx}
\usepackage{multirow}
\usepackage{color}
\usepackage{array}
\usepackage{natbib}

\usepackage{xcolor}
\definecolor{RoyalBlue}{HTML}{4169e1}
\definecolor{ForestGreen}{HTML}{228b22}

\usepackage[colorlinks = true,
            linkcolor = blue,
            urlcolor  = blue,
            citecolor = blue,
            anchorcolor = blue]{hyperref}

\usepackage{comment}

\newcommand{\arsham}[1]{{\color{black} #1}}

\begin{document}
\title{Toward a Physics of Deep Learning and Brains}

\author{Arsham Ghavasieh}
\thanks{Corresponding author.}
\email{arsham.ghavasieh@gmail.com}
\affiliation{Center for Complex Networks and Systems Research, Luddy School of Informatics, Computing, and Engineering, Indiana University, Bloomington, Indiana 47408, USA}

\author{Meritxell Vila-Mi\~nana}
\thanks{These authors contributed equally to this work.}
\affiliation{Center for Complex Networks and Systems Research, Luddy School of Informatics, Computing, and Engineering, Indiana University, Bloomington, Indiana 47408, USA}

\author{Akanksha Khurd}
\thanks{These authors contributed equally to this work.}
\affiliation{Center for Complex Networks and Systems Research, Luddy School of Informatics, Computing, and Engineering, Indiana University, Bloomington, Indiana 47408, USA}

\author{John Beggs}
\affiliation{Department of Physics, Indiana University, Bloomington, Indiana 47405, USA}

\author{Gerardo Ortiz}
\affiliation{Department of Physics, Indiana University, Bloomington, Indiana 47405, USA}
\affiliation{Institute for Advanced Study, Princeton, NJ 08540, USA}

\author{Santo Fortunato}
\thanks{Corresponding author.}
\email{santo@iu.edu}
\affiliation{Center for Complex Networks and Systems Research, Luddy School of Informatics, Computing, and Engineering, Indiana University, Bloomington, Indiana 47408, USA}

\date{\today}

% ---------- ABSTRACT ----------
\begin{abstract}
Deep neural networks and brains both learn and share superficial similarities: processing nodes are likened to neurons and adjustable weights are likened to modifiable synapses. But can a unified theoretical framework be found to underlie them both? Here we show that the equations used to describe neuronal avalanches in living brains can also be applied to cascades of activity in deep neural networks. These equations are derived from non-equilibrium statistical physics and show that deep neural networks learn best when poised between absorbing and active phases. Because these networks are strongly driven by inputs, however, they do not operate at a true critical point but within a quasi-critical regime--- one that still approximately satisfies crackling noise scaling relations. By training networks with different initializations, we show that maximal susceptibility is a more reliable predictor of learning than proximity to the critical point itself. This provides a blueprint for engineering improved network performance. Finally, using finite-size scaling we identify distinct universality classes, including Barkhausen noise and directed percolation. This theoretical framework demonstrates that universal features are shared by both biological and artificial neural networks.
\end{abstract}

\maketitle

\begin{figure}
    \centering
    \includegraphics[width=\linewidth]{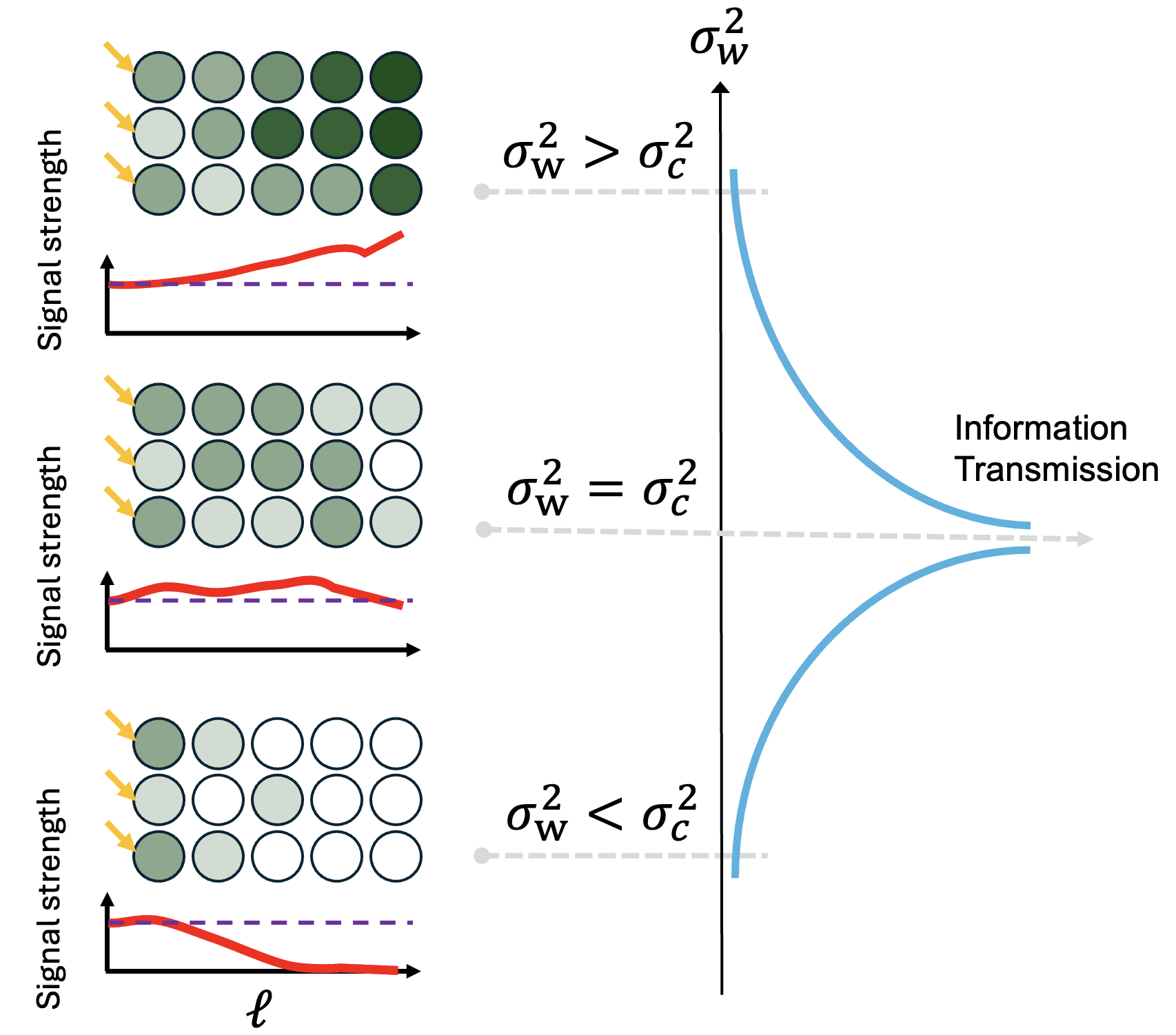}
    \caption{\textbf{Avalanches and criticality.} Schematic representation of signal propagation through layers $\ell$ in deep neural networks. Circles are the neurons, columns of circles constitute layers of deep network, the darker they get the more active they are (higher $y^{\ell+1}$, defined in the text). Signal strength (denoted by $\sqrt{q_{\ell}}$ in the text) characterizes the size of the neural gains ($z^\ell$) or, also called pre-activations. Let $\sigma_w^2$ be a control parameter, with $\sigma_c^2$ being its critical value. When $\sigma_w^2 > \sigma_c^2$ signals grow with depth, when $\sigma_w^2 < \sigma_c^2$ they decay, and when in the thermodynamic limit $\sigma_w^2 = \sigma_c^2$, signals remain stable, enabling ideal information propagation. Neural networks can best preserve and process information with initializations of the weight parameters near the critical regime $\sigma_w^2 \approx \sigma_c^2$. The initial signal strength defines the avalanche threshold. The cumulative signal strength above the threshold value--- the area between red line and purple dashes--- gives the avalanche size, $S$, and the number of layers it penetrates until it crosses the threshold defines the avalanche's duration $D$.}
    \label{fig:emblematic}
\end{figure}

Biological neuronal systems have long been studied through statistical physics. Maximum-entropy and equilibrium-like models provide a powerful lens through which to study memory storage \cite{hopfield1982neural, cossart2003attractor} and criticality \cite{schneidman2006weak, tkavcik2015thermodynamics}. However, living neural networks produce directed cascades of activity, suggesting an event based non-equilibrium approach. The concept of neuronal avalanches--- i.e., spatiotemporal bursts of activity separated by silent periods--- provides such an account that matches scale-free statistics observed in neuronal data, with dynamic fluctuations typical of critical phase transitions \cite{beggs2003neuronal}. Yet the original version of this framework required that external inputs to the network would be small and rarely occur. A more realistic formulation is often referred to as quasi-criticality\cite{qcriticality2014, eqcriticality2021} in contrast with plain criticality. It states that as neuronal populations are driven, they hover within a tunable neighborhood of a critical point\cite{qcriticality2014, eqcriticality2021}.

But are living neural networks actually quasi-critical\cite{qcriticality2014}? Early accounts relied on global proxies such as the branching ratio to argue that neuronal systems operate near a critical point (see \cite{Cortex-Criticality2022} for a review). By contrast, drawing on advances in non-equilibrium statistical physics, current studies verify proximity to criticality using the battery of tests provided by crackling noise theory \cite{sethna2001crackling, zapperi2022crackling}. These include the power law distribution of avalanche sizes $P(S) \sim S^{-\tau_{s}}$ and durations $P(D) \sim D^{-\tau_{d}}$, the scaling of average size with duration $\langle S \rangle_D \sim D^{\gamma}$, culminating in an exponent (scaling) relation $\frac{\tau_d - 1}{\tau_s - 1}\approx \gamma$ extensively used to quantify distance from criticality\cite{ma2019cortical}, together with the universal shape collapse of rescaled avalanche profiles \cite{friedman2012universal,ponce2018whole,qcriticality2014,eqcriticality2021}.

Why, from an evolutionary standpoint, should these networks operate near criticality\cite{Cortex-Criticality2022}? Computational systems can take a variety of advantages by operating near a critical point, including maximum dynamic range \cite{kinouchi2006optimal, shew2009neuronal}, information transmission\cite{greenfield2001mutual, shew2011information} and computational power\cite{bertschinger2004edge}. %input sensitivity and responsiveness without saturation. 
From an information-theoretic view, proximity to criticality elevates susceptibility and Fisher information, providing steep input-output gain and enhanced stimuli discriminability\cite{livi2017determination}.

Deep learning has traced a parallel arc to neuroscience. Perceptrons~\cite{Rosenblatt1958}, the ground-breaking first trainable neural models, were initially limited to single-layer computations. Near equilibrium, energy-based networks such as Hopfield models~\cite{hopfield1982neural} and Boltzmann machines~\cite{ACKLEY1985} enabled learning with hidden representations whose recent generalizations enable a much larger memory capacity~\cite{DenseAM_krotov}. A dynamical account of deeper networks suggested that performance hinges on proximity to a critical line, or the edge of chaos, separating ordered and chaotic regimes~\cite{poole2016exponential,schoenholz2016deep, Yang2017MeanFieldResNet, Zhang2021EdgeOfChaosGuiding,Tamai2023UniversalScalingEdge,Feng2019OptimalIntelligenceEdge,day2025feature}. While being at the edge of chaos does not generally guarantee performance--- and our work provides a novel explanation for that--- the edge of chaos framework has nonetheless illuminated the behavior of very deep architectures. \arsham{Recent work further suggests that, over successive generations, deep architectures have converged toward criticality \cite{vock2025critical}, as reflected in vanishing Lyapunov exponents and branching ratios approaching $1$. A natural next step is to look for a more explicit non-equilibrium physics formulation to answer whether deep networks benefit from scale-free event statistics, obey scaling laws, face constraints from finite-size effects, and belong to identifiable universality classes.}

Decades of progress in deep learning now permit more fundamental questions about the relationship between criticality and distributed (neural) computation. Deep learning can, as neuroscience has, leverage advances in nonequilibrium physics to sharpen its characterization of criticality, since it presently shares several of that field’s early limitations. Firstly, most evidence of criticality in deep networks still relies on architecture-level proxies like finite-time Lyapunov exponents rather than on the statistics of events. Secondly, it is important to consider that deep networks are strongly driven. Inputs typically perturb large fractions of the first layer. As mentioned before, large drive pushes systems away from exact criticality, suggesting that quasi-criticality provides a more relevant organizing principle \cite{qcriticality2014}. Thirdly, crackling noise theory offers a unified framework with concrete predictions for the interrelations of observables near the critical point. It can be used to test whether deep learning actually occurs near a critical phase transition. Finally, it is worth mentioning that not all criticality is alike. In fact, distinct regimes corresponding to different universality classes are plausible with direct consequences for functionality, that are currently indistinguishable.

In this work, we perform an event-resolved, crackling-noise analysis for deep learning. First, in Gaussian-initialised networks we characterise deep avalanches, show that their size and duration distributions are power laws, obtain their exponents, verify the crackling noise scaling relation between the exponents, and find that avalanche shapes collapse onto a universal curve to establish a genuine non-equilibrium phase transition of the Barkhausen universality class. Second, we link computation to dynamics by showing that trainability overlaps a quasi-critical plateau: the control parameters giving the heightened susceptibility region also enable learning. Third, analyzing three ResNet~\cite{he2016deep} variants, we find critical dynamics again but with exponent relations consistent with mean-field directed percolation. Together, these results move beyond global proxies like the Lyapunov exponents by providing an event resolved approach that unravels for the universality of learning in deep networks, required for practical diagnostics for locating and steering models within quasi-critical regimes.

\section{Deep network dynamics.} 

\arsham{In this work, we focus on how the dynamical regime at initialization allows trainability by providing dynamical flexibility and responsiveness. We do not generally expect trained networks to remain near criticality, since tasks introduce characteristic scales and drive dynamics away from criticality, motivating recent efforts to identify critical subspaces rather than globally critical post-training dynamics~\cite{fontenele2024lowdimensional}.}

Consider a feedforward network of depth $L$ and uniform layer width $N$, which is the number of neurons per layer. Let $W^{\ell}_{ij}$ be the weight of the connection from neuron $j$ in layer $\ell - 1$ to neuron $i$ in layer $\ell$. Let $b^{\ell}_{i}$ be the bias of neuron $i$ in layer $\ell$. 

The weights and biases are Gaussian distributed: $W^{\ell}_{ij} \sim \mathcal{N}\Big(0,\frac{\sigma_w^2}{N}\Big)$ and $b^\ell_i \sim \mathcal{N}(0,\sigma_b^2)$, with zero mean and variances of $\frac{\sigma_w^2}{N}$ and $\sigma_b^2$, respectively. 
Neuron $i$ on layer $\ell$ has activity $y_i^{\ell+1}$, resulting from applying an activation function $\phi$ on a weighted sum of preceding layer activities plus the neuron's bias, 
\begin{equation}
y_i^{\ell+1} = \phi\big(z_i^{\ell}\big), 
\label{eq:forward-activation}
\end{equation}
where the gain function (also called pre-activations) follows
\begin{equation}\label{eq:pre-activation}
    z_i^{\ell}=\sum_{j} W^{\ell}_{ij}\, y_j^{\ell}+b^\ell_i.
\end{equation}
Here, we use $\phi = \tanh$ as our activation function, unless stated otherwise.

In the next section, we use a mean-field treatment to show that Gaussian deep networks undergo a dynamical phase transition.

\section{Mean-field approximation to dynamics.}

Here we use a mean-field theory developed to study the evolution of neural gains (Eq.~\ref{eq:pre-activation})~\cite{schoenholz2016deep}.
In the mean-field limit $N \to \infty$, the central limit theorem ensures that
the gains become Gaussian random variables $\mathcal{N}(0,q^{(MF)}_{\ell})$, fully characterized by their variance $q^{(MF)}_{\ell} = \mathbb{E}[(z_i^\ell)^2]$.  We denote the variance steady state as $\lim\limits_{\ell \to \infty} q_{\ell}^{(MF)}=q_{ss}^{(MF)}$ and show that 

\begin{eqnarray}
    q^{(MF)}_{ss}&\sim& \Big(\sigma_w^2 - 1 \Big)^\beta, \quad   \sigma_w^2\to1^{+}, \; \sigma_b^2 =0 \\
    q^{(MF)}_{ss}&\sim& \Big( \sigma_b^2\Big)^{\beta/\sigma}, \qquad   \sigma_w^2=0 ,\;\;\;\; \sigma_b^2 \to 0^{+}
\end{eqnarray}
with $\beta = 1$ and $\sigma = 2$. These exponents are consistent with the mean-field directed percolation (MF DP) universality class~\cite{Lbeck2005}. For the derivations, see Appendices.

Among the response functions that diverge at the critical point, we focus on the \emph{$\sigma_w$-susceptibility} to characterize the sensitivity of signal strength ($\sqrt{q^{MF}_{ss}}$) to fluctuations in connectivity $\sigma_w^2$:
\begin{eqnarray}\label{eq:susceptibility}
    \chi^{(MF) }_{\sigma_w^2} &=& \frac{d}{d\sigma_w^2}\sqrt{q^{(MF)}_{ss}} 
\;\sim \;\Big(\sigma_w^2-1\Big)^{-1/2} \nonumber \\ \sigma_w^2 \to && 1^{+}, \;\;\; \sigma_b^2 = 0,
\end{eqnarray}
\emph{$\sigma_w$-susceptibility} directly relates neural gains to learning which is primarily achieved through connectivity alterations. For more information, see Appendices.

It is important to note that another characterization of criticality exists for Gaussian deep networks, based on order and chaos, that we discuss in the next section.

\section{Edge of chaos and performance.}

Here, we introduce the cross-input correlations and how they unravel the \textit{edge of chaos} in Gaussian deep networks, as a framework widely used to understand why some deep networks learn better than others~\cite{schoenholz2016deep}.

Let $z_{i;a}^{\ell}$ denote the neural gains (pre-activation) specifically under input $a$. The layerwise cross-input covariance $q^{ab}_{\ell}$ and correlation $C^{ab}_{\ell}$, are defined as
\begin{eqnarray}
q^{ab}_{\ell} &=& \mathbb{E}\left[z_{i;a}^{\ell}\, z_{i;b}^{\ell}\right]
= \frac{1}{N}\sum_{i=1}^{N} z_{i;a}^{\ell} z_{i;b}^{\ell} \\
C^{ab}_{\ell} &=& \frac{q^{ab}_{\ell} }{\sqrt{q^{aa}_{\ell} q^{bb}_{\ell} }}
\label{eq:q_ab}
\end{eqnarray}

Let $ \lim\limits_{\ell\to\infty}C^{ab}_{\ell}=C^{ab}_{ss}$ be the steady-state cross-input correlation. Let us rewrite the correlation at layer $\ell$ as $C^{ab}_{\ell} = C^{ab}_{ss} + \delta_{\ell}^{ab}$, where $\delta_{\ell}^{ab}$ is the deviation from steady state. Through mean-field analysis, it has been shown~\cite{schoenholz2016deep} that the deviation evolves like $\delta_{\ell}^{ab} = e^{-\frac{\ell}{\zeta_c}}$, where $\zeta_c$ is the \textit{cross-input correlation depth}. It reveals the edge of chaos, i.e, a curve in the ($\sigma_w^2,\sigma_b^2$) plane along which the cross-input correlation depth $\zeta_c$ diverges, $\zeta_c\to\infty$. In other words, correlations between distinct inputs remain essentially unchanged as they traverse the networks whose ($\sigma_w^2,\sigma_b^2$) lie on this curve. 

This has been used as a framework to explain why some network initializations are not trainable for large depths $L$~\cite{schoenholz2016deep}. Certainly, the edge of chaos provides insights into performance differences across parameter choices. However, proximity to the edge of chaos does not guarantee trainability~\cite{schoenholz2016deep}. In fact, except for very small $\sigma_b^2$, the points on the line are poorly trainable, raising an important challenge. Does it mean that criticality cannot explain learning performance? To provide an answer, we first compare the edge of chaos with a Widom-like line in the next section.

\begin{figure*}
    \centering
    \includegraphics[width=\linewidth]{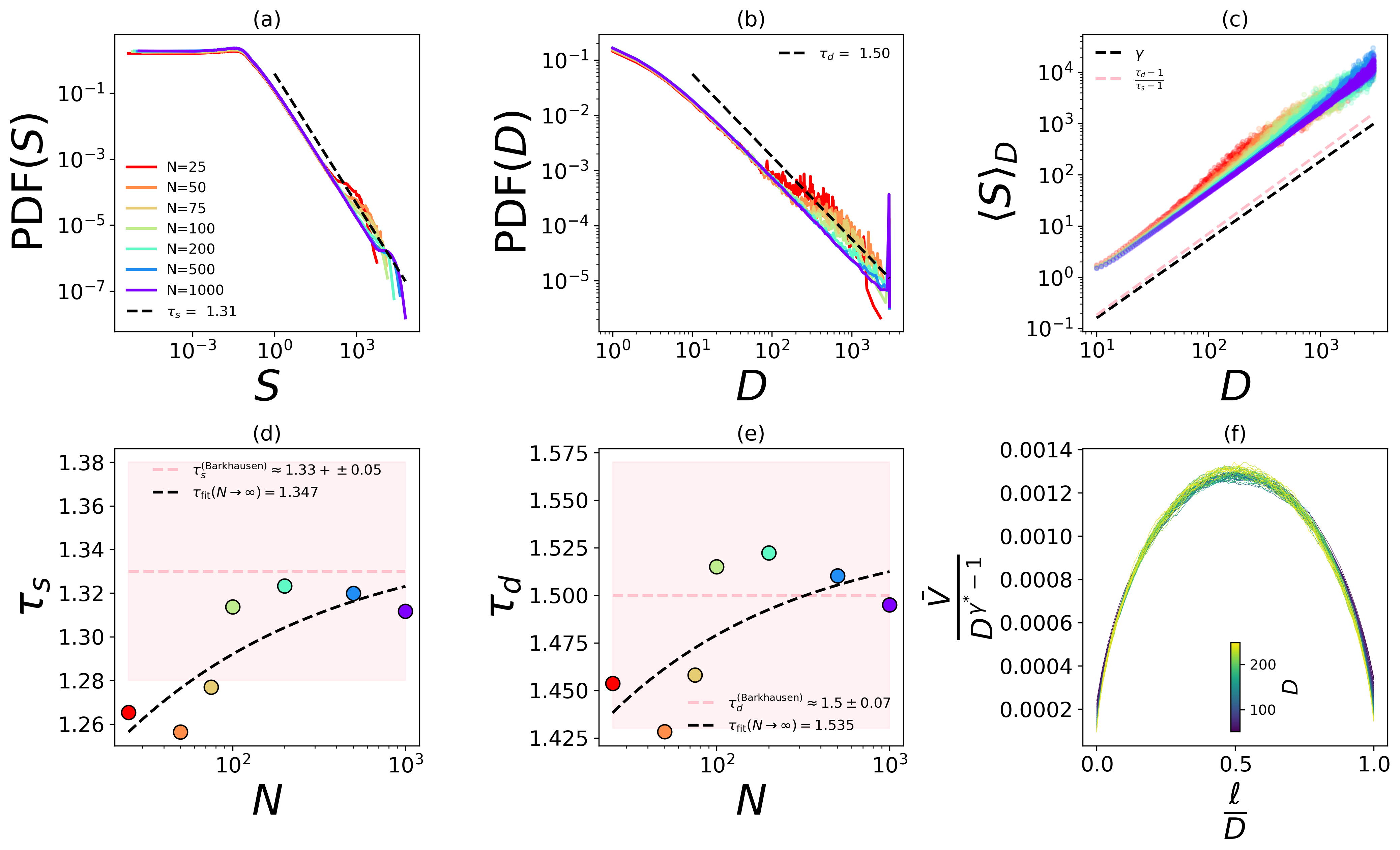}
    \caption{\textbf{Crackling noise statistics in Gaussian initialised deep networks.} (a-b) Distributions of avalanche size $S$ and duration $D$ for fixed depth $L=3000$ and varying widths $N=25, 50, 75, 100, 200, 500, 1000$, with zero bias $\sigma_b^2=0$ and near-critical weights $\sigma_w^2\approx1$. Networks are re-initialised every $5000$ perturbations until $\sim4\times10^6$ avalanches are collected (See Text). Maximum-likelihood fits are used to find the exponents. (c) Size–duration scaling: $\langle S\rangle_D\sim D^{\gamma}$ for different widths with $\gamma\approx 1.53$ being the slope fitted to $N=1000$--- close to the theoretical prediction $\frac{\tau_d - 1}{\tau_s - 1}\approx 1.58$. (d-e) Finite-size extrapolation of exponents via $\tau(N)=\tau+cN^{-w}$ gives $(\tau_s,\tau_d)\to(1.34,1.53)$ for large $N$. Pink dashes show Barkhausen values with their errorbars as shades $(1.33\pm0.05,\,1.50\pm0.07)$. (f) Shape collapse analysis with optimal rescaling value $\gamma^{\ast}\approx 1.58$, very close to the prediction $\frac{\tau_d - 1}{\tau_s - 1}\approx 1.58$ (See Text and Appendices for more information).}
    \label{fig:avalanche_stats}
\end{figure*}

\section{Edge of chaos vs Widom-like line.}

Here, we analyze and contrast two competing notions of optimality: one defined by a critical boundary separating ordered and chaotic regimes, and the other characterized by maximal dynamical susceptibility arising from fluctuations in connectivity.

Criticality at the edge of chaos is about cross-input correlations (Eq.~\ref{eq:q_ab}). Important though this is, it does not fully characterize deep propagation. Even when cross-input correlations are preserved, signal amplitudes associated with each input can relax rapidly to a steady state, bleaching information~\cite{schoenholz2016deep}. This motivates a complementary question: is there another critical line along which sensitivity to inputs, or responsiveness, diverges?

The mean-field critical behavior we observed for deep propagation belongs to the directed percolation universality class. Therefore, another type of criticality can be studied, where response functions like $\sigma_w$-susceptibility diverges (at $\sigma_b^2=0)$ --- showing that the dynamics is maximally sensitive to fluctuations in the weights of the network links. 

However, unlike the edge of chaos, we expect that increasing $\sigma_b^2$ erases the non-analyticity associated to this type of criticality. Such expectation has physically meaningful roots. Mechanistically, each neuron’s activity reflects two contributions: couplings from the preceding layer $\sum\limits_{j} W_{ij}y_j$ and its bias $b_i$. By analogy with condensed matter physics and neuroscience, biases act as external fields or spontaneous activity, tilting responses and breaking explicitly the symmetry required for a sharp critical transition. 

Therefore, for $\sigma_b^2 > 0$, susceptibility is expected to exhibit finite peaks around $\sigma_w^2 \approx 1$--- both in mean-field calculations and simulations. The locus of these maxima define the Widom-like line in the ($\sigma_w^2, \sigma_b^2$) plane.

It is important to note that the edge of chaos and the Widom-like line intersect only at the point of exact criticality $(\sigma_w^2,\sigma_b^2)=(1,0)$ and depart for $\sigma_b^2>0$, while Widom-like line's peaks shrink until it eventually vanishes. \arsham{We show that the Widom-like line determines the onset of trainability and its fading correlates with the width of the trainable region in the phase space (See Fig.~\ref{fig:susc_tracks_trainabilitiy} and Fig.~\ref{fig:Widom_line}).}

In the next section we go beyond the mean-field approximations to provide an event-based understanding of deep networks. If a system is genuinely at a critical phase, these events are expected to follow specific power law distributions, with exponents approximately satisfying the theoretical predictions of crackling noise theory~\cite{sethna2001crackling,zapperi2022crackling}.

\section{Deep avalanches.}
Avalanches are spatiotemporal cascades of activity bounded by periods of silence. They characterize a variety of physical phenomena, from snow and land slides, earthquakes, fractures and cracks to flux lines in type II superconductors, biological neurons and brain areas~\cite{zapperi2022crackling}. However, they have not been previously characterized in deep neural networks. Here, we obtain them from the signal strength $\sqrt{q_\ell}$ evolution.

\begin{figure*}
    \centering
    \includegraphics[width=\linewidth]{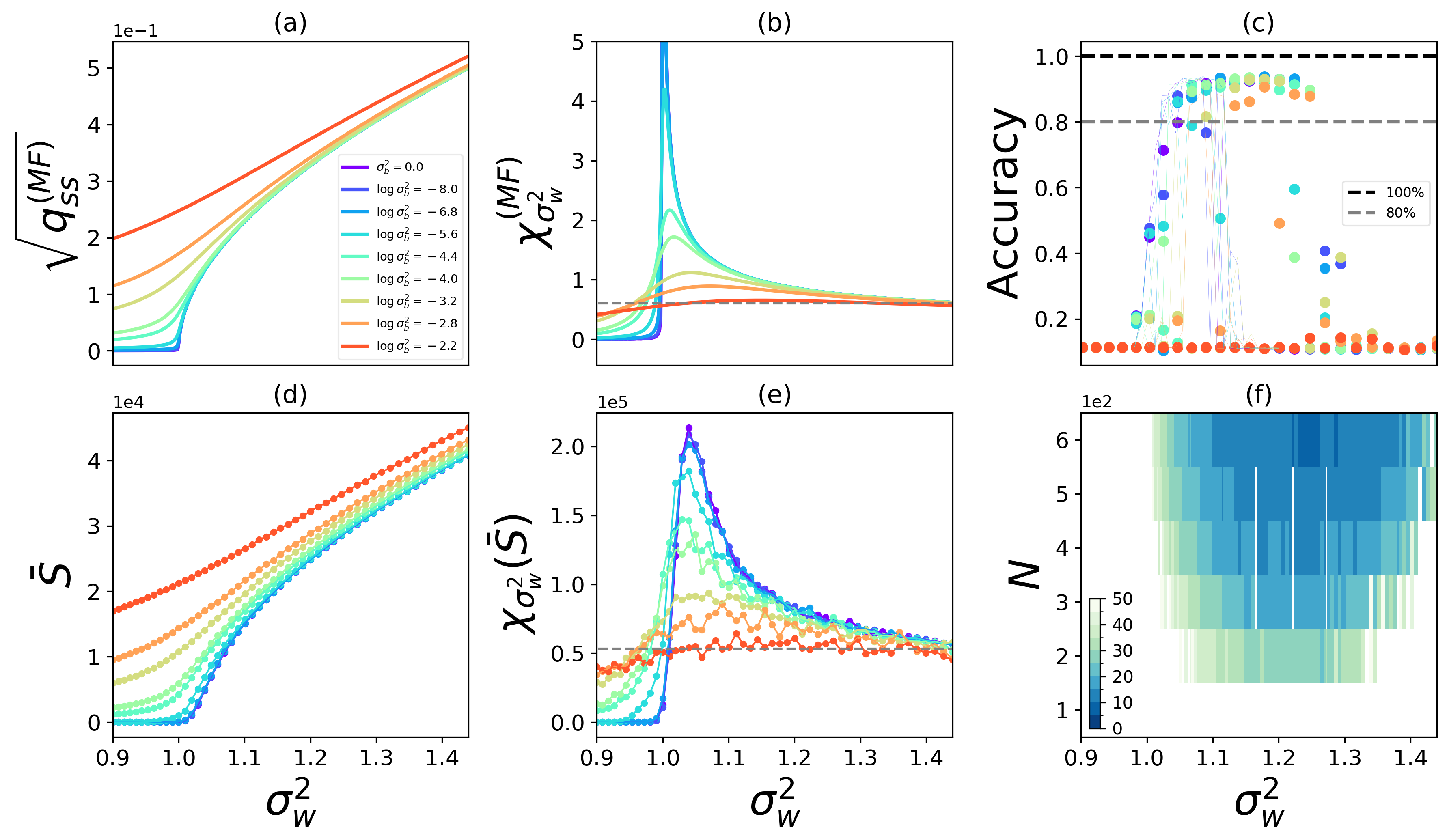}
    \caption{\textbf{Mean-field predictions and simulations of susceptibility, and learning performance.} Color indicates bias variance: all curves in panels (a–e) are color-coded according to $\log_{10} \sigma_b^2$ as shown in the legend in panel (a), ranging from $\sigma_b^2 = 0$ (purple) to $\sigma_b^2 = 10^{-2.2}$ (red). 
    (a) Steady-state signal strength $\sqrt{q^{(MF)}_{ss}}$ from mean-field theory. The logarithm bases are 10. 
    (b) Mean-field $\sigma_w$-susceptibility $\chi^{(MF)}_{\sigma_w^2}$. 
    (c) Test performance on the MNIST classification task after 10 epochs. 
    (d) Empirical average avalanche size $\bar S$, which is the sum of all recorded avalanche sizes divided by the number of recorded avalanches. 
    (e) Empirical $\sigma_w$-susceptibility (from avalanche statistics $\chi_{\sigma_w^2} = \frac{d\bar S}{d\sigma_w^2}$). Panels (c–e) are measured on networks with width $N=300$ and depth $L=300$. 
    (f) Epochs needed to reach 97\% training accuracy for fully connected deep neural networks with $L=300$, as a function of network width $N$ and fixed $\sigma_b^2=0$. White regions indicate models that did not reach 97\% within 50 epochs. \arsham{The onset of successful learning approaches $1^+$ as size increases, which is expected for a finite system approaching its thermodynamic-limit transition.}}
    \label{fig:susceptibility}
\end{figure*}

Based on our diverging mean-field susceptibility (Eq.~\ref{eq:susceptibility}), we expect to observe a transition in the system's behavior at $(\sigma_w^2, \sigma_b^2)\approx (1,0)$, reflected in the evolution of signal strength $\sqrt{q_\ell}$. As illustrated in Fig.~\ref{fig:emblematic}), the input signal strength $\sqrt{q_0}$ may be attenuated, amplified, or remain approximately constant as it traverses the network. These three regimes mirror subcritical, supercritical, and critical dynamics, respectively, with the control parameter $\sigma_w^2$ determining the transition between them at its critical value $\sigma_c^2\approx 1$.

Guided by this intuition, we define avalanches relative to the input signal strength $\sqrt{q_0}$, which we take as the threshold. The number of layers in which $\sqrt{q_\ell}$ remains above the threshold $\sqrt{q_0}$ specifies the avalanche duration $D$, while the cumulative signal strength across these layers yields the avalanche size $S \propto \sum\limits_{\ell=1}^{D}\sqrt{q_\ell}-\sqrt{q_0}$. For a detailed definition, see Appendices.

Near a genuine critical point, avalanches are expected to exhibit scale-free statistics, with size and duration distributions following power laws, $P(S)\sim S^{-\tau_s}$ and $P(D)\sim D^{-\tau_d}$. Also, it is expected that their exponents are not independent. They must satisfy the crackling-noise scaling relation
$\gamma \approx \frac{\tau_d - 1}{\tau_s - 1}$~\cite{sethna2001crackling, zapperi2022crackling}, where $\gamma$ characterizes the scaling behavior
$\langle S \rangle_D \;\sim\; D^{\gamma}$, with $\langle S \rangle_D$ denoting the average size of avalanches of duration $D$. Beyond these scaling relations, a hallmark of critical avalanches is their universal temporal shape. When individual avalanche profiles are rescaled by their duration $D$ along the layer axis and by $D^{\gamma-1}$ in amplitudes, they are expected to collapse onto a single curve.~\cite{sethna2001crackling, zapperi2022crackling}. See Appendices for mathematical details.

\section{Criticality in Gaussian initialized deep networks.} 

Here we explore the avalanche statistics of Gaussian deep networks, demonstrating that they satisfy what is theoretically expected from systems near a critical point.

In Fig.~\ref{fig:avalanche_stats}, we consider Gaussian initialized models of fixed depth $L=3000$ and varying widths $N=25, 50, 75, 100, 200, 500, 1000$ with no bias $\sigma_b^2 = 0$ and near critical weights $\sigma_w^2 \approx 1$. In fact, the weight variances are selected a little larger than the exact critical value $\sigma_w^2(N) - 1 = \delta\sigma_w^2(N) > 0$ to compensate for finite-size effects. In ascending order of widths, the correction values obtained by scanning for maximal straightness of the distributions in the log–log plots over a range of possible correction values are $\delta\sigma_w(N) = 0.03, 0.015, 0.009, 0.005, 0.002, 0.001, 0.001 $, reported to three decimal places and ordered by increasing $N$. To make an avalanche, we sample Gaussian inputs with size $\sqrt{q_0} = 0.1$ (See Appendices) and perturb the first layer of the network with. Each network is reinitialized every $5000$ perturbations (weights and biases are resampled from their Gaussian distributions) to reduce the sampling noise. We stop when approximately four million valid avalanches are obtained. 

In Fig.~\ref{fig:avalanche_stats}-(a-b) we show that the distributions of avalanche size $S$ and duration $D$ are power laws spanning more than three and two decades, respectively (For more information on plots and fits, see Appendices). The fixed input $q_{0} = 0.01$ is used for all widths, resulting in an input of $0.01/N$ per neuron, shifting the curves in a size-dependent way. We resolve the issue by setting $S\rightarrow \sqrt{N} S$, that well aligns the starting points of size distributions as shown in Fig.~\ref{fig:avalanche_stats} (a). We use maximum likelihood to find the best power laws with their corresponding exponents $S^{-\tau_s}, D^{-\tau_d}$ (See Appendices). Also, we show that the average size of avalanches with the same duration $\langle S\rangle_D$ scales well with their duration: $\langle S \rangle_D \sim D^{\gamma}$ (Fig.~\ref{fig:avalanche_stats}-(a-b)). 

For the largest system, $N=1000$, we report the fitted slope in $\langle S \rangle_D \sim D^{\gamma}$ to be $\gamma = 1.5303$ and the optimal value for the shape collapse (Fig.~\ref{fig:avalanche_stats}-(f)) to be $\gamma^* = 1.5800 \pm 0.1131$, both being in proximity of the theoretical prediction $\frac{\tau_d - 1}{\tau_s - 1} = 1.5883$. We plot the size and duration exponents with respect to the layer widths and fit a line $\tau(N) = \tau + c N^{-w}$ to it, where $\tau = \tau(N = \infty)$ estimates the exponents in the limit of infinite system size (Fig.~\ref{fig:avalanche_stats}-(d-e))--- a standard way to remove finite size effects. Our results show that $(\tau_s, \tau_d) \rightarrow (1.34, 1.53) $ at large $N$. Models have found Barkhausen noise exponents of $(1.34,1.55)$~\cite{Cerruti2006} and experimental values like $(1.33 \pm 0.05 , 1.5 \pm 0.07)$\cite{bohn2018playing}-- error bars of Barkhausen noise exponents in Fig.~\ref{fig:avalanche_stats} --, both within a reasonable distance from our values.  

In this section the main hallmarks of criticality~\cite{zapperi2022crackling} have been shown in deep networks. But how does criticality relate with learning and performance? We answer that in the next section, primarily using $\sigma_w$-susceptibility (Eq.~\ref{eq:susceptibility}).

\section{Learning and quasi-criticality.}\label{sec: subsection learning and quasi-criticality}

Here, we provide a comprehensive analysis of the phase transition in signal strength $\sqrt{q_\ell}$ and $\sigma_w$-susceptibility through mean-field theory and avalanche simulations, ultimately exploring how they relate to learning performance (See Fig.~\ref{fig:susceptibility}).

The mean-field analysis maps how the stationary signal strength 
$\sqrt{q^{(MF)}_{ss}}$ and the susceptibility vary across the $(\sigma_w^2,\sigma_b^2)$ parameter space. Consistent with the external field analogy, increasing
$\sigma_b^2$ raises $\sqrt{q^{(MF)}_{ss}}$ and rounds the transition, eliminating the true critical point. Consequently, susceptibility develops a sharp ridge of large response, also called Widom-like line, near $\sigma_w^2\approx1,\; \sigma_b^2 \approx 0$. 

Note that every point on the edge of chaos is “critical” only in the sense of a diverging cross-input correlation depth. By contrast, the magnitude of the $\sigma_w$-susceptibility along the Widom-like line diminishes \arsham{for larger $\sigma_b^2$.} If learning depends on signal penetration depth rather than cross-input correlation depth alone, trainability should decline at large $\sigma_b^2$ as the Widom ridge flattens and ultimately dissolves.

We test these predictions in a finite network with width $N=300$ and depth $L=300$. Simulations align with mean-field theory in terms of the mean avalanche size $\langle S\rangle$, the steady-state signal strength $\sqrt{q^{(MF)}_{ss}}$, and the $\sigma_w$-susceptibility $\chi_{\sigma_w^2}$. Learning performance follows the same landscape: accuracy after 10 epochs peaks in the region of heightened susceptibility. Moreover, as $\sigma_b^2$ increases, the trainable region narrows and shifts to larger $\sigma_w^2$, mirroring the susceptibility ridge.

\begin{figure*}
    \centering
    \includegraphics[width=\linewidth]{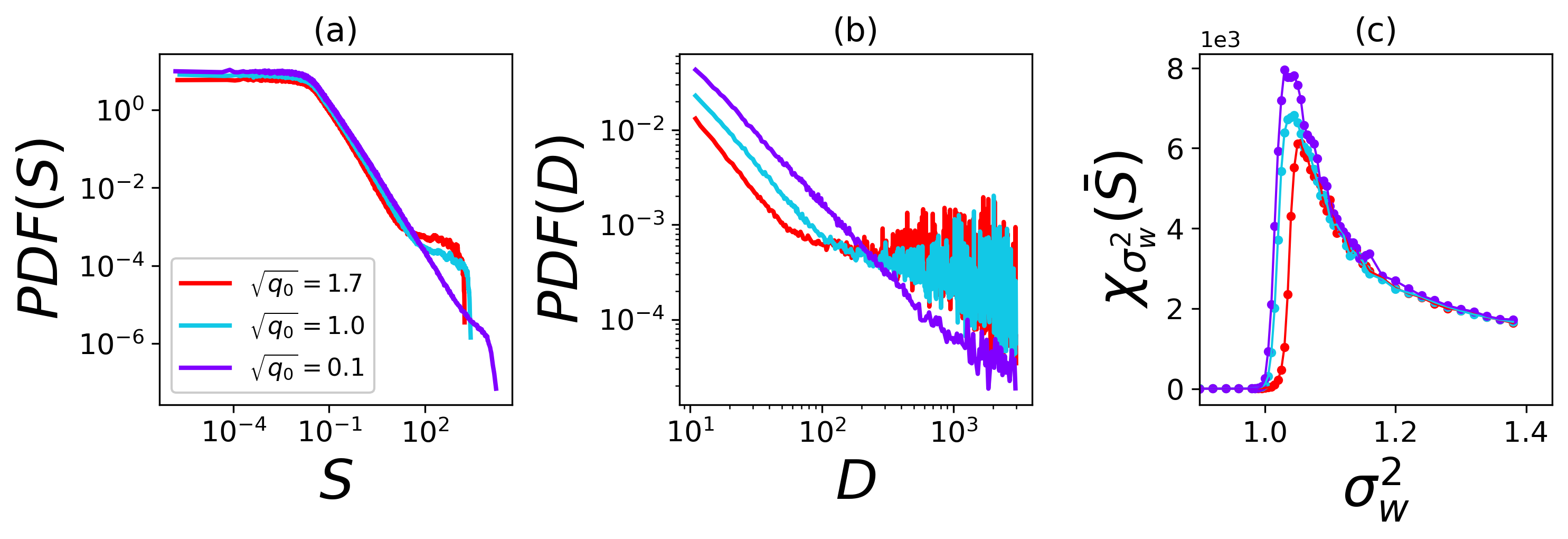}
    \caption{\textbf{Input magnitude and quasi-criticality.} (a) avalanche size distribution, (b) duration distribution, and (c) $\sigma_w$-susceptibility curves for varying input magnitudes $\sqrt{q_0}$. The maximum in the susceptibility shifts to larger $\sigma_w^2$ values as the input  magnitude increases from $\sqrt{q_0}=0.1$ to $\sqrt{q_0}=1.7$ at the point where the $\sigma_w$-susceptibility rises sharply (c). Consequently,  better power laws of (a) and (b) for input magnitudes $\sqrt{q_0}=0.1, 1, 1.7$ are, respectively, obtained at $\sigma_w^2 \approx 1.005, 1.025, 1.05$.}
    \label{fig:input_criticality}
\end{figure*}

In Fig.~\ref{fig:susceptibility}, we measure the learning performance using the MNIST digit classification task\cite{lecun1998mnist}. The training accuracy is the fraction of correctly classified digits. We plot the training accuracy reached after 10 epochs on the MNIST classification task for different $(\sigma_w^2, \sigma_b^2)$ initialization pairs, using a fully connected deep neural network with a depth of $300$ layers, and $300$ neurons per hidden layer. We observe that successful training overlaps with the area of heightened susceptibility. As $\sigma_b^2$ grows, the area moves to the right side, including larger values of $\sigma_w^2$, and its width shrinks. At $\sigma_b^2 =10^{-2.2}$, the network is not trainable under this task, reflecting the low $\sigma_w$-susceptibility. These findings suggest that learning tasks require certain levels of proximity to criticality. 

We also check the effect of network width in the case of $\sigma_b^2 = 0$. Specifically, we train fully connected deep neural networks with a fixed depth of $L=300$ layers and with varying hidden layer widths of 200, 300, 400, 500, and 600 neurons per layer, initializing weights and biases as in Fig.~\ref{fig:susceptibility}-(c). In contrast to the previous experiments, here we fixed the bias variance to $\sigma_b^2=0$ and varied the weight variance $\sigma_w^2$ within a narrow range around the critical regime. Each network was trained for at most 50 epochs, with early stopping applied if a training accuracy of 97\% was achieved earlier. In Fig.~\ref{fig:susceptibility}-(f) we report the number of epochs needed to reach 97\% accuracy in each case. Further details regarding the network structure, task, and experimental setup are provided in the Appendices section Training and quasi-criticality on MNIST dataset \ref{ref: subsection MNIST methods}. The results show that trainability and, thus, learnability starts to be achieved for those networks initialized near the critical value $\sigma_w^2 = 1$, highlighting the dependency of the network's performance on the initialization. The width of the trainable area increases with the network's width. Similar performance areas, indicated with the same color shades in Fig.~\ref{fig:susceptibility}, occur  closer and closer to $\sigma_w^2 \approx 1$, as $N$ grows, confirming the expectation of finite size effects. A detailed comparison of the training accuracy achieved for different depths (200-500 layers) is provided in Supplementary Information, showing that being near to criticality is even more vital for deeper networks.

\arsham{We extend the analysis to a more challenging dataset, CIFAR-10, and obtain comparable results that further confirm the relationship between criticality and trainability (See Appendices).

In fact, we find extremely large Pearson correlations between the loci of susceptibility peaks and the onset of trainability ($98\%$ for MNIST and $99\%$ for CIFAR-10), as well as considerable Pearson correlations between the logarithmic susceptibility height and the width of the trainable region ($83\%$ for MNIST and $58\%$ for CIFAR-10). Given the high correlations, it is natural to think that the width of the trainable envelope is a function of susceptibility height, and a clean fit is of course possible, but here we suffice to a logarithmic relation. See Fig.~\ref{fig:susc_tracks_trainabilitiy}.

\begin{figure}
    \centering
    \includegraphics[width=\linewidth]{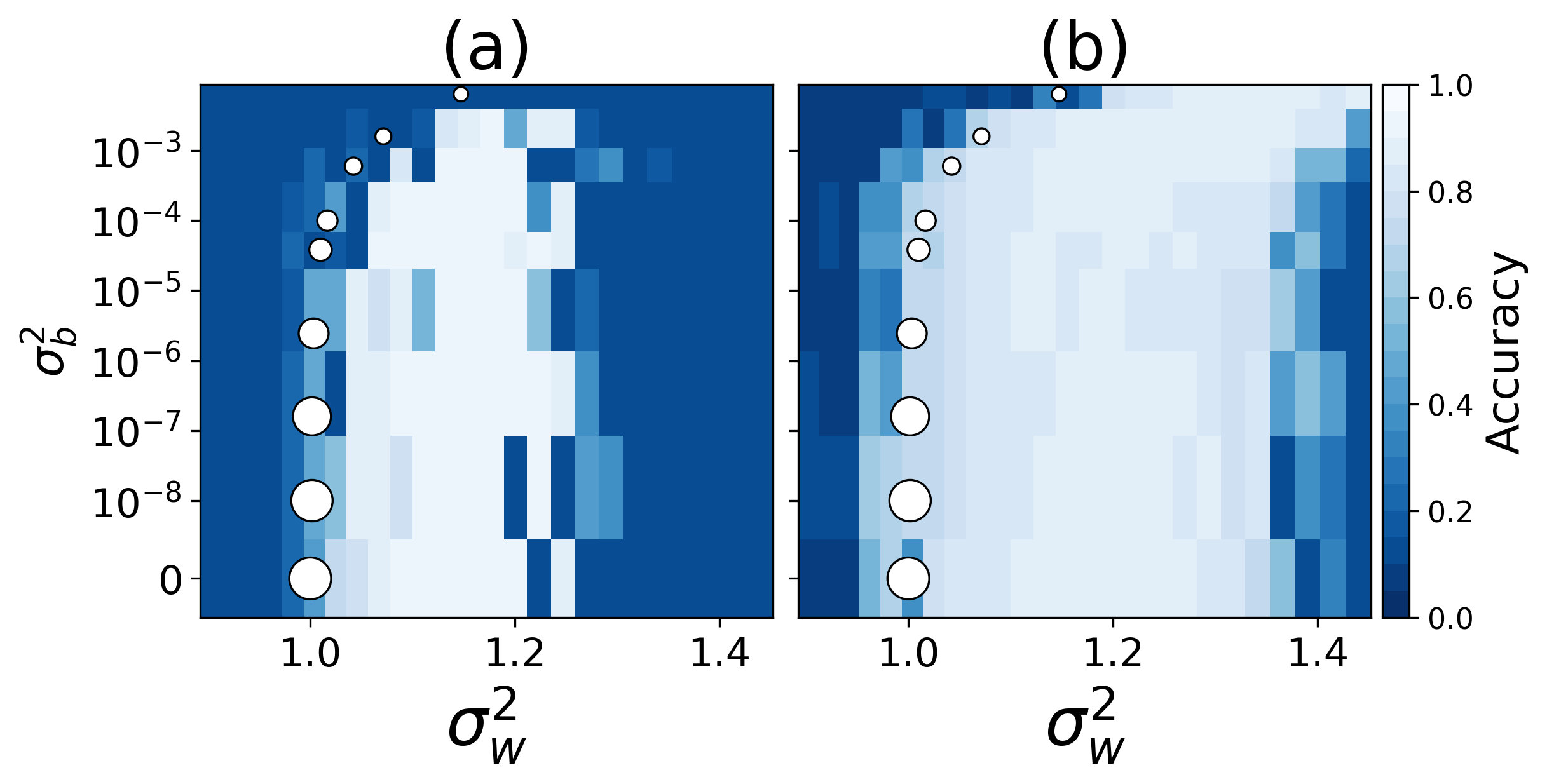}
    \caption{\arsham{\textbf{Trainability and susceptibility across initialization parameters.}
    Heat maps show the training accuracy for MNIST (a) and CIFAR-10 (b) across the same $(\sigma_w^2,\sigma_b^2)$ control-parameter space. Circles mark the locations of susceptibility peaks, with circle size proportional to the peak height. In both datasets, the susceptibility peaks track the onset of trainability, linking heightened dynamical responsiveness to the region where optimization becomes successful.}}
\label{fig:susc_tracks_trainabilitiy}
\end{figure}

}

In  models of brain dynamics, the probability $p_s$ that a neuron is spontaneously activated accounts for intrinsic noise or external input~\cite{qcriticality2014}. As $p_s$ increases, neurons activate more frequently, potentially disrupting the separation of time scales between the driving and relaxation processes. As a result, the peak of the dynamical susceptibility shifts within the parameter space, and the effective exponents and cutoffs of the size and duration distributions are altered. In a sense, $p_s$ plays a role analogous to an external field $h$ in magnetic systems or to $\sigma_b^2$ in our deep neural network formulation, as explained earlier. However, in deep neural networks the input magnitude $\sqrt{q_0}$ is another important parameter that  lacks a clear counterpart in the neuroscience literature. Similarly to $p_s$ and $\sigma_b^2$,  the input magnitude also  activates neurons, but it influences only the first layer and the initial stage of the process. Because this effect is locally and temporally confined, it does not affect the relaxation dynamics or the separation of time scales.  Figure~\ref{fig:input_criticality} shows the effect of $\sqrt{q_0}$ on avalanches and $\sigma_w$-susceptibility profiles. Interestingly, the input magnitude has little effect on the effective exponents, yet it clearly shifts the system away from the critical point, reducing the susceptibility peak and altering its position. Note that the input data are often mean-centered ($x\to x - \mathbb{E}[x]$) and then standardized ($x \to x/\sqrt{\mathbb{E}[x^2]}$) so that their typical magnitude, quantified by the root-mean-square value, is approximately $\sqrt{q_0}=1$. This offers another reason why deep neural networks, at least those with Gaussian initialization, cannot be exactly critical, but are better described in terms of quasi-criticality.

\section{ResNets}

While we have shown that the signatures of criticality are strong in Gaussian deep networks, the field of artificial intelligence has developed a wide array of architectures designed to perform specific task. In contrast with the Gaussian networks where the structures emerge through learning out of a blank slate like initialization, these architectures are highly engineered. In the next section, we study three famous deep convolutional architectures to show that they, too, exhibit signatures of criticality.

\begin{figure*}
    \centering
    \includegraphics[width=\linewidth]{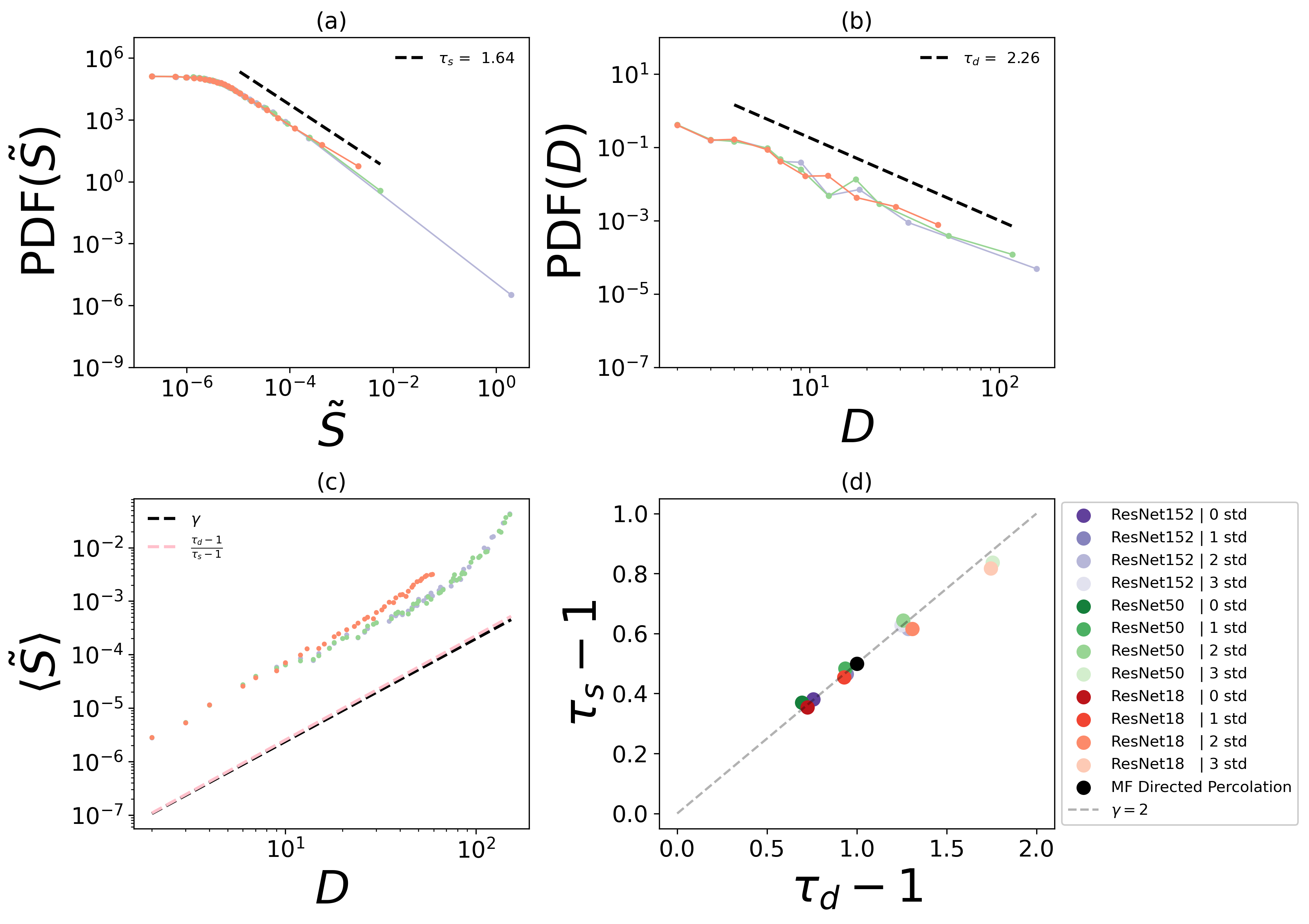}
    \caption{\textbf{Avalanche statistics in ResNets exhibit crackling-noise scaling} (a) The distributions of avalanche size $\tilde{S}$ (See Text and Appendices) and (b) duration $D$ follow power laws. The scaling of average avalanche size and duration (c) for multiple ResNet variants thresholded with $n=2$, or two standard deviations above the mean activity of the layer (see Appendices), suggests crackling noise relationship. (d)  The exponents ($\tau_s,\tau_d$) are closely clustered regardless of the thresholding parameter $n$. $\gamma \approx \frac{\tau_d - 1}{\tau_s - 1} \approx 2$ indicates operation near a critical point, well aligned with the universality class of  mean-field directed percolation.}
    \label{fig:resnet}
\end{figure*}

Unlike the Gaussian-initialized networks, in which logical operations mostly emerge through learning and self-organization, Residual Networks (ResNets)~\cite{he2016deep} are highly engineered systems with precise sequences of operations including convolutions, normalization, nonlinearities, and identity skip connections. Yet, recent works suggest that they, too, exhibit signatures of criticality, making them an interesting empirical case to study through the lens of crackling noise theory~\cite{vock2025critical}. Technically, the structure of ResNet calls for an addendum in our previous definitions of avalanche properties, details of which can be found in Appendices. We denote the avalanche sizes defined for ResNet by $\tilde{S}$, to distinguish from the previous sections.

Figure~\ref{fig:resnet} shows that the distributions of avalanches in ResNets follow power laws. However the power laws and the crackling scaling are not as strong as Gaussian initialized networks. Whether it shows that there is room for improvement in terms of avalanche measurement, or in designing convolutional networks with clearer power-law statistics, is unclear and requires future investigation. However, it is worth mentioning that in ResNets, the modules like the BatchNorm get tuned to the data statistics in the training phase, which is not reflected in our analysis as we exclude training. While our current work is focused on deep networks at initialization, it might be possible that training iterations prevent the runaways we observe for large avalanches (See Fig.~\ref{fig:resnet}-(c)).

Notably, ResNet versions are not the same architectures at varying sizes. They have different designs--- operations and sequences. Yet the size and duration exponents stay very close to each other for different thresholdings $n = 0, 1, 2, 3$, suggesting robust crackling noise scaling. This indicates that ResNets actually operate near a critical point and confirms other recent works~\cite{vock2025critical}. In addition to this proximity, we obtain $\gamma \approx 2$ regardless of thresholding--- the line with $\gamma = 2$ in the $\tau_s, \tau_d$ plot aligns with the mean-field directed percolation universality class. 

\begin{figure}
    \centering
    \includegraphics[width=0.8\linewidth]{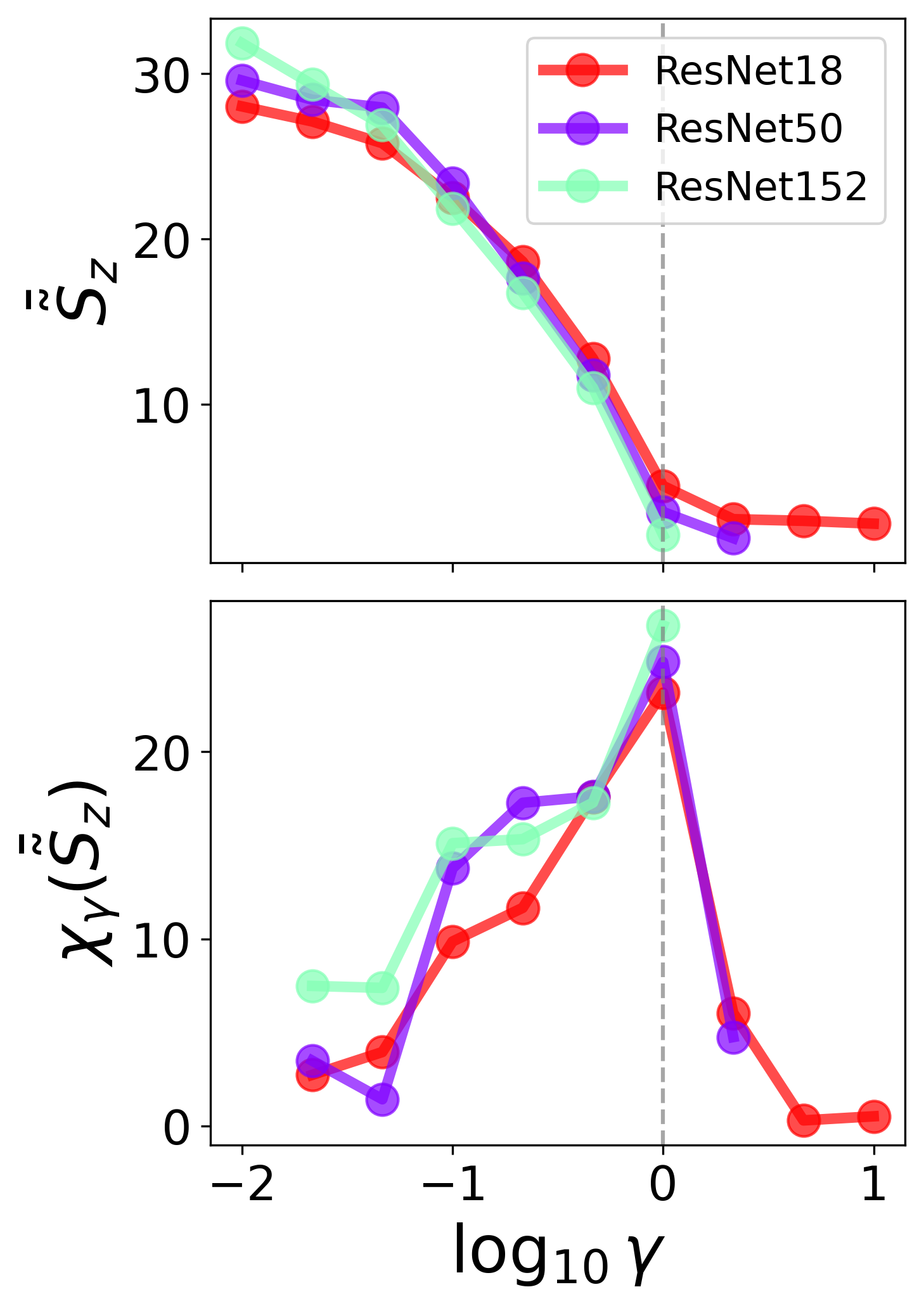}
    \caption{\arsham{\textbf{Critical transition in ResNets controlled by the BatchNorm scale parameter.}
    (a) The order parameter $\langle \tilde{S}_z\rangle$ (see Eq.\ref{eq:relative_fluctuation_size}) as a function of $\log_{10}\gamma$ for ResNet18, ResNet50, and ResNet152. Varying the BatchNorm affine scale $\gamma$ drives a sharp change in network activity near $\gamma=1$ (vertical dashed line). 
    (b) The corresponding susceptibility $\chi_{\gamma}= \frac{\Delta\bar{\tilde{S}}_z}{\Delta \log_{10}\gamma} $ peaks near $\gamma=1$, identifying the critical region. Remarkably, the susceptibility peak lies at the onset of the successful training range, while performance degrades for $\gamma>1$ and for sufficiently small $\gamma$~\cite{Davis2022Revisiting}.}}
    \label{fig:resnet_susceptibility}
\end{figure}

\arsham{Comparing avalanches across ResNet configurations initialized with different parameters is difficult because both the mean activity and its standard deviation, which determine the threshold used to identify avalanches, vary substantially across configurations. We therefore first characterize the typical fluctuation size of each configuration. Denoting the standard deviation of the activity at layer $\ell$ by $\sigma_\ell$, we approximate the total fluctuation scale as

\begin{eqnarray}
\sigma_{\mathrm{total}} \sim \sum_\ell \sigma_\ell ,
\end{eqnarray}

where cross-layer correlations are neglected. We then normalize each fluctuation size by this characteristic fluctuation scale and define

\begin{eqnarray}\label{eq:relative_fluctuation_size}
\tilde{S}_z = \frac{\sum\limits_\ell \sqrt{q_\ell}}{\sigma_{\mathrm{total}}} .
\end{eqnarray}

Thus, $\tilde{S}_z$ proxies avalanche size relative to the typical fluctuation size of each network. We use its average over the ensemble as an order parameter that can be compared across ResNet configurations. For each choice of parameters, we draw $500$ network initializations and record $100$ runs for each, yielding $5\times 10^4$ measurements of $\tilde{S}_z$. Its mean over the ensemble $\bar{\tilde{S}_z}$ gives the order parameter, and the BatchNorm affine parameter $\gamma$ acts as a control parameter, driving the system toward a critical phase transition at $\gamma \simeq 1$, where the corresponding susceptibility peaks (See Fig~\ref{fig:resnet_susceptibility}). The transition is more clearly resolved when $\gamma$ is represented on a logarithmic scale. 

Interestingly, this boundary is aligned with previous systematic analyses reporting degraded performance for $\gamma>1$ and for very small values near $\gamma<0.05$, whereas values between approximately $0.1$ and $1$ achieve strong performance depending on the learning rate. The observation of a critical transition near $\gamma\simeq1$, marking the onset of trainability, therefore provides independent support for our theoretical prediction.}

Overall, this section shows that the hallmarks of criticality can be found not only in Gaussian initialized networks, but also in highly engineered deep structures like ResNets.

\section{Discussion}

Taken together, our work reveals a link between crackling noise theory, artificial intelligence and living brains. We provided an event-resolved, non-equilibrium phase transition framework to understand deep learning. By resolving propagation into avalanches we validated crackling noise predictions like power-law distributed sizes and durations, their mutual scaling, the exponent relationship $\gamma \approx \frac{\tau_d - 1}{\tau_s - 1}$ measuring distance from criticality, and the universal shape collapse revealing self-similar propagation of information. Remarkably,
these predictions have also been identified in biological neural networks~\cite{Cortex-Criticality2022}. In addition, we identify distinct universality classes for Gaussian initialized deep networks and ResNets, respectively, matching Barkhausen and mean-field directed percolation classes.

\begin{figure}
    \centering
    \includegraphics[width=\linewidth]{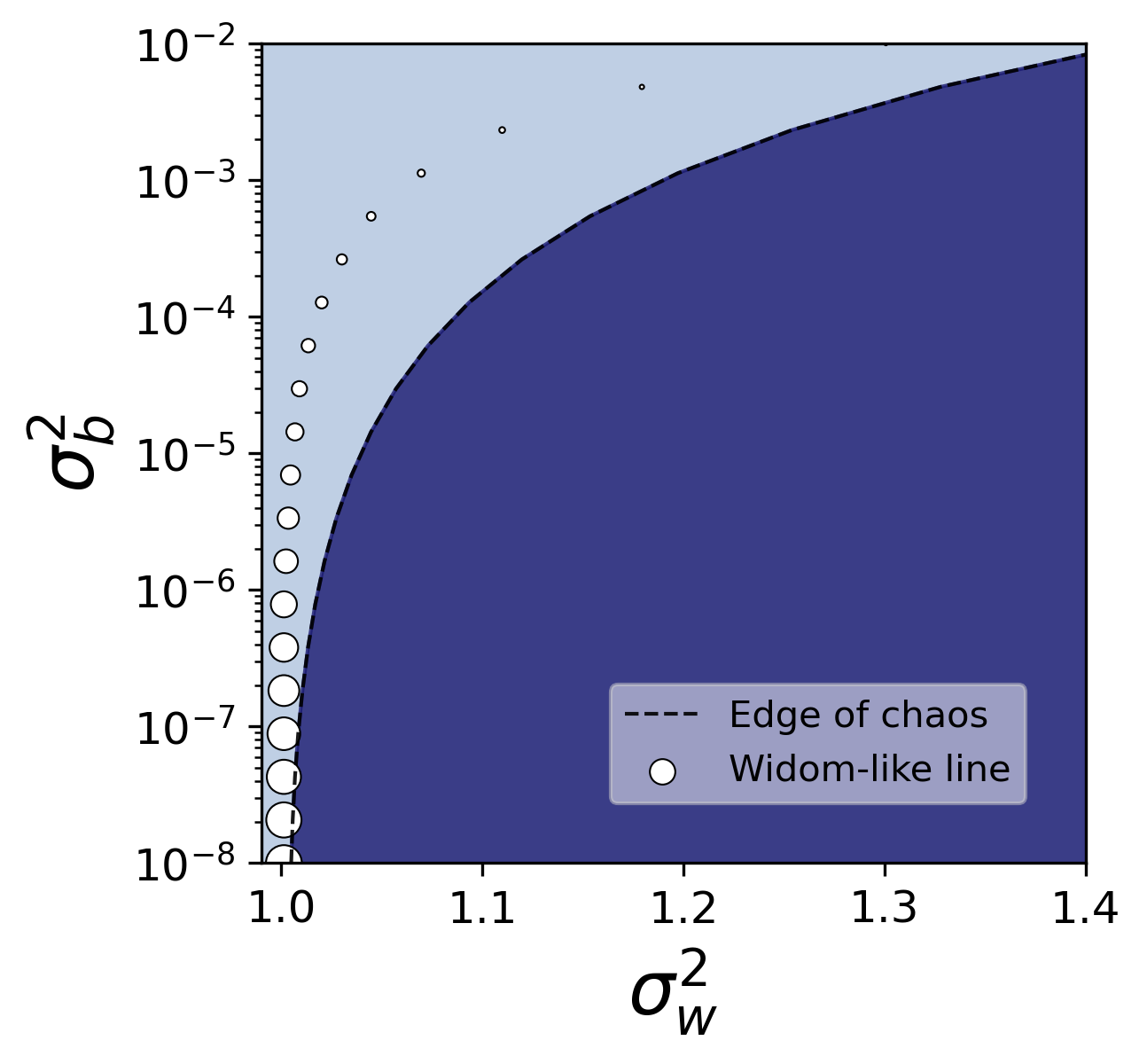}
    \caption{\textbf{Mean-field Widom-like line and the edge of chaos.} The edge of chaos marks parameters where the cross-input correlation depth diverges--- with light and dark blue indicating the ordered and chaotic regions, respectively. In contrast, the Widom-like line locates the maximum \emph{$\sigma_w$-susceptibility} where sensitivity to inputs is largest--- the size of the dots encode the height of maximum susceptibility. The two curves meet only at the $(\sigma_w^2,\sigma_b^2)=(1,0)$. Importantly, all points on the edge of chaos correspond to the divergence of cross-input correlation depth. However, as $\sigma_b^2$ grows, the susceptibility exhibits weaker peaks until the Widom-like line completely vanishes--- well aligned with the learning performance (See Text).}
    \label{fig:Widom_line}
\end{figure}

Criticality is often taken to guarantee computation~\cite{kinouchi2006optimal,shew2009neuronal, greenfield2001mutual, shew2011information}. However, networks on the edge of chaos can train poorly if the bias is non-negligible. We explain this long-standing puzzle in a quasi-criticality framework\cite{qcriticality2014, eqcriticality2021}: biases act as external fields, destroying the critical point when judged by susceptibility rather than cross-input correlations. The only critical point where penetration diverges is $(\sigma_w^2,\sigma_b^2) = (1,0)$--- which also dissolves for large inputs. Instead, a Widom-like line of maximal but finite susceptibility replaces the critical point for $\sigma_b ^2> 0$ (See Fig.~\ref{fig:Widom_line}). Empirically, learning aligns with this quasi-critical plateau along the Widom-like line, and not with the entirety of the edge of chaos.

Our findings have practical implications. The crackling toolkit provides operational diagnostics for training. Pipelines can be developed for tracking the exponent-relation mismatch, collapse quality, and, ultimately, distance from criticality. Standard knobs like dropout, spectral constraints, batch normalization, injected noise and residual depth can serve as more advanced control parameters for criticality--- for instance, in terms of their effect on the susceptibility. Accordingly, regularizers can bias training toward the quasi-critical plateau, and phase diagrams that guide architecture and hyper-parameter selection. Moreover, we predict that the task specific performance can vary between different criticalities--- characterized by sets of exponents $\{ \tau_s, \tau_d, \gamma\}$. Steering universality classes can potentially lead the design of a new generation of deep architectures. 

Our study also has its limitations. Finite system sizes and subsampling impose cut-offs that complicate exponent estimation. We addressed this with maximum-likelihood fits and finite-size scaling analysis, yet larger-scale analysis can improve the confidence. While we tracked the pre-activations in Gaussian networks, and the produced tensors by modules in ResNet, there are a variety of observables that might show signatures of criticality but go beyond the scope of this work. They include but are not limited to the post-activations in Gaussian networks, block level propagations in ResNet, the slopes of susceptibility and signal strength curves through mean-field approximations and many-body simulations. Finally, being limited to Gaussian networks and ResNet architectures, our work invites broader tests of quasi-criticality in deep learning.

In conclusion, we provided a novel framework based on crackling noise theory for artificial intelligence, showing that deep networks can indeed operate near criticality, that criticality can predict performance, and that learning is supported not by exact point criticality but by quasi-critical plateaus. This shared physics with neuroscience~\cite{qcriticality2014,eqcriticality2021} offers both mechanistic insight and a design playbook for building and steering future generation models. Both brains and deep neural networks use avalanches or cascades of activity to transmit information through stages or layers of processing units. For information to be preserved in this architecture, the final layer must receive activity that is neither attenuated nor saturated. Operating near the critical point best satisfies this requirement. Because both brains and deep neural networks are strongly driven by inputs, though, they must operate in a quasi-critical regime where they are {\it as critical as possible.} These commonalities are not merely superficial, but fundamental, and lead both systems to share physical laws describing maximal susceptibility along the Widom line. Elucidating these laws is a first step toward building a common physics for deep learning and brains.

\section*{Author contributions} A. G. designed the project and performed most calculations, analytical and numerical. M. V. M. and A. K. performed calculations on deep learning performance. J. B., G. O. and S. F. provided feedback on the various phases of the project. All authors wrote the paper. 

\section*{Acknowledgments}
The authors thank Minsuk Kim, 
Robert Jankowski, Bendeguz Sulyok and Emilio Cobanera for their feedback during the first stages of the project. JB was supported by a subaward from NSF Expedition “Mind in Vitro” award \#IIS–2123781.  G.O. gratefully acknowledges support from the Institute for Advanced Study. 

\section*{Code Availability}
The open-source code is available on GitHub at  \url{https://github.com/ArshamCS/DNN_criticality.git}.
\section*{Competing Interests Statement}
The authors declare no competing interests.	

\bibliographystyle{apsrev4-2}
\bibliography{references} % references.bib file

\appendix

\section{Mean-field $\sigma_w$-susceptibility.}

As weights and biases are independent, the signal energy is $q^{(MF)}_{\ell} = \sigma_w^2 \mathbb{E}\left[\big(y^{\ell-1}\big)^2\right] +\sigma_b^2$. Since $y^{\ell} = \phi(z^{\ell-1})$ and $z^{\ell-1}\sim\mathcal{N}(0,q_{\ell-1}^{(              MF)})$, we get a recursion: $q^{(MF)}_{\ell} = \sigma_w^2 \int Dz \phi^2\big(\sqrt{q^{(MF)}_{\ell-1}}\,z\big) +\sigma_b^2$,
where $Dz = \frac{dz}{\sqrt{2\pi}}e^{-z^2/2}$. In the limit of $\ell \to \infty$, we obtain the steady state equation
\begin{equation}
q^{(MF)}_{ss} = \sigma_w^2 \int Dz \phi^2\big(\sqrt{q^{(MF)}_{ss}}z\big) +\sigma_b^2.
\label{eq:steady_state}
\end{equation}

We show that $(\sigma_w^2,\sigma_b^2)=(1,0)$ is a candidate for a critical point of a continuous phase transition. Expanding $\phi^{2}(x) \approx x^2 - \tfrac{2}{3}x^4 $ using Gaussian moments $\mathbb{E}[z^2]=1$, $\mathbb{E}[z^4]=3$, we have  
$q^{(MF)}_{ss}\,\Big[(1-\sigma_w^2) + 2\sigma_w^2 q^{(MF)}_{ss}\Big] = 0$, at $\sigma_b^2 = 0$. Besides the trivial solution for the signal strength $q^{(MF)}_{ss}=0$, the nontrivial branch $q^{(MF)}_{ss} = \frac{1}{2} \Big(1-\frac{1}{\sigma_w^2}\Big)$. 
For $\sigma_w^2 - 1 \ll 1$, in the steady state $q^{(MF)}_{ss}\sim \Big(\sigma_w^2 - 1 \Big)^\beta, \quad \beta =1$ at $\sigma_w^2 \approx 1$ and $ \sigma_b^2=0$. Similarly, for $\sigma_b^2>0$ and  $\sigma_w^2=1$, after expanding again we obtain $q^{(MF)}_{ss} = q^{(MF)}_{ss}\Big[\sigma_w^2  - 2\sigma_w^2 q^{(MF)}_{ss}\Big] + \sigma_b^2$, leading to $q^{(MF)}_{ss} = \frac{1}{2}\Big(\sigma_b^{2}\Big)^{1/2}\sim \Big( \sigma_b^2\Big)^{\beta/\sigma}$, with $\sigma = 2$. The exponents $\beta = 1$ and $\sigma = 2$ are the ones of the universality class of mean-field directed percolation~\cite{Lbeck2005}, reflecting the same critical behavior.

The variable $\sqrt{q^{(MF)}_{ss}}$ is the \textit{signal strength}.
The steady state pre-activations $z_{ss}$ form a vector of magnitude $\sqrt{q^{(MF)}_{ss}}$ rather than $q^{(MF)}_{ss}$. The signal strength  $\sqrt{q^{(MF)}_{ss}}$ determines the mean-field evolution, as it enters the gain function $\phi\Big(\sqrt{q^{(MF)}_{ss}}z\Big)$. Moreover, in deep propagation, the amplification or attenuation of signals is better reflected in the signal strength than its square $q^{(MF)}_{ss}$. Therefore, the \emph{$\sigma_w$-susceptibility} is defined based on the signal strength:
\begin{eqnarray}
\chi^{(MF) }_{\sigma_w^2} &=& 
\frac{d}{d\sigma_w^2}\sqrt{q^{(MF)}_{ss}} \nonumber \\
&=& \frac{1}{2\sqrt{2}\,\sigma_w^3\,\sqrt{\sigma_w^2-1}} \nonumber \\ &\sim& \Big(\sigma_w^2-1\Big)^{-1/2} \nonumber \\ \sigma_w^2 &\to& 1^{+}, \sigma_b^2 = 0.
\end{eqnarray}
The $\sigma_w$-susceptibility is 
a response function diverging at the critical point, that captures the behavior of the signal strength (See Fig.~\ref{fig:emblematic}). 

Similarly, it can be shown that the \emph{$\sigma_b$-susceptibility} $\chi_{\sigma_b^2}^{(MF)} \; = \; \frac{d\sqrt{q^{(MF)}_{ss}}}{d\sigma_b^2} \; \sim \; \Big(\sigma_b^{2}\Big)^{-\frac{3}{4}}$ diverges at the critical point--- as $\sigma_b^2 \to 0$.
However, here we primarily examine the \emph{$\sigma_w$-susceptibility}, since it quantifies how response diversity varies with fluctuations in the link weights of the network ($\sigma_w^2$), a factor directly relevant for learning.

\section{Avalanche characterization.}
Here we detail the mathematics of avalanches in Gaussian intialized deep networks.

Tracking avalanches is often simpler in discrete dynamical systems. For instance, in case of spiking neurons, an avalanche starts with a spike. If at least one neighboring neuron spikes in response, the avalanche continues to propagate. On the contrary, the avalanche dies when the causal chain stops--- with no firing from neighbors of those neurons that fired in the previous step. The avalanche duration is the time passed from its start to end. The size of the avalanche is the number of spikes it contains.

In continuous systems, like the whole brain electrical activity, crossing a threshold defines when an avalanche starts and ends~\cite{qchumans2022}. The input strength $\sqrt{q_{0}}$ constitutes our threshold. This is suggested by what normally happens at a critical point, where the strength of the signal does not decay or amplify.  It also aligns well with the discrete systems scenario, where the avalanche activity never goes below its starting point--- 1 active neuron.

In our experiments, we fix the input strength $\sqrt{q_{0}}$, and track the response of the network. If the signal strength goes below the input strength for the first time  at layer $\ell_{f}$ (i.e., $\sqrt{q_{\ell_{f}}}<\sqrt{q_{0}}$), the avalanche ends at layer $\ell_{f}$. The duration (or depth) of the avalanche is $D = \ell_{f} - 1$. Note that the step at which input enters the system ($\ell =0$) and the last step where the strength goes below the threshold ($\ell = \ell_f$) are not included in the duration. Also, the avalanche size is $S  = \sqrt{N}\times\Big(\sum\limits_{\ell = 1}^{\ell_f-1} \sqrt{q_{\ell}} - \sqrt{q_{0}}\Big)$. Note that $q_{\ell}$ is an average over neurons in layer $\ell$. Therefore, $\sqrt{q_\ell}$ is normalized by $\sqrt{N}$, and the factor $\sqrt{N}$ in the avalanche definition turns the mean into the total strength.

We sample the input at the beginning of each avalanche from a Gaussian distribution $z^{0} = \mathcal{N}(0, 1)$. Then, we impose the desired input strength $\sqrt{q_{0}}$ by performing the transformation of input:
$z^0 \rightarrow \sqrt{q_{0}}  \frac{z^0}{\|z^0\|}$, ultimately guaranteeing $\| z^0 \| = \sqrt{q_{0}}$ . We use $q_{0}=0.01$ and record the avalanche sizes, durations and shapes for further analyses.

It is worth mentioning that the constant threshold $\sqrt{q_{0}}$ we use for Gaussian initialized networks of uniform widths has limited applicability. For instance, it can not work well with the convolutional networks where width and operations are highly variable from layer to layer. In that case, we use a layer dependent threshold.

\section{Distributions, exponents, and fitting.} To visualize the power-law behavior of strength and duration for a given network configuration, we plot their probability density functions (PDFs) on log–log scales using an adaptive binning approach. The data (either strength or duration) is first sorted in ascending order, and the smallest value defines the left edge of the first bin. Consecutive bin edges are then chosen such that each bin contains a minimum of k data points, until the data is exhausted. This approach yields narrow bins in dense regions (near the lower end of the distribution) and wider bins in sparser regions (the heavy tail). The counts for each bin are normalized by the bin width and the total number of observations or total area under the curve for discrete and continuous data, respectively. Finally, the center of each bin is calculated as the geometric mean of its edges. The power-law fits to extract the exponents $(\tau_s, \tau_d)$ were implemented using the \texttt{powerlaw} library in python \cite{alstott2014powerlaw}.

As mentioned in the text, the estimate of $\gamma$ can be obtained from the scaling of mean avalanche size with duration $\langle S \rangle_D = D ^ {\gamma}$. Taking logarithms gives, we plot the points with $\log\langle S \rangle_D$
and $\log{D}$. We use a weighted least squares regression to fit the line $
\log \langle S \rangle_D 
\approx \gamma \, \log D + b$, where each duration $D$ is assigned a weight $
w(D) = \frac{N_D}{\mathrm{Var}[\log S]_D}$, with $N_D$ being the number of avalanches of duration $D$ and 
$\mathrm{Var}[\log S]_D$ the sample variance of their logarithmic sizes (For this purpose, we omit the D values for which only one avalanche is recorded). This choice reduces the importance of durations with fewer avalanches or noisier statistics.   

\section{Shape-collapse analysis.}

An alternative way to determine $\gamma$ is through the shape-collapse analysis. In addition to that, shape collapse analysis unravels the scale-free property of events at the microscopic level.

The evolution of $\sqrt{N}\Big(\sqrt{q_{\ell}}-\sqrt{q_0}\Big)$ from the start of an avalanche $\ell=1$ till its end $\ell = D$ defines the avalanche profile. For the $i-$th recorded avalanche of duration $D$ where the signal strength at layer $\ell$ is denoted by $\sqrt{q^{(i,D)}_\ell}$, let $V_D^i(\ell)=\sqrt{N}\Big(\sqrt{q_{\ell}^{(i,D)}}-\sqrt{q_0}\Big)$ denote the avalanche profile. For $N_D$ avalanches of duration $D$, we find the mean avalanche shape as
\begin{equation}
\bar V_D(\ell)=\frac{1}{N_D}\sum_{i=1}^{N_D} V_D^i(\ell).
\end{equation}
We rescale $\ell$ as $u = \frac{\ell}{D}$
so that all avalanches are defined in the unit interval. To have the same number of points in all avalanche shapes, each mean shape $\bar V_D(u)$ is then recalculated onto a common grid of $n$ points,
$ u_j = \frac{j+\tfrac{1}{2}}{n_{\rm grid}}, 
\; j = 0,1,\dots,n-1$,
using linear interpolation to obtain the values $\bar V_D(u_j)$.

For a choice of $\gamma$, each mean profile is rescaled by $D^{\gamma-1}$ as implied by scaling theory~\cite{zapperi2022crackling}:
\begin{equation}
\tilde V_D(u_j;\gamma)=\frac{\bar V_D(u_j)}{D^{\gamma-1}},
\end{equation}
which is expected to be independent of $D$ for the right choice of $\gamma$ at the critical point. In other words, $\tilde V_D(u_j;\gamma)$ is expected to reveal the fractal geometry of critical propagations. The shape collapse analysis assesses that.

Assume the minimum and maximum durations considered in the analysis be $D_{min}$ and $D_{max}$, respectively. The mean value of the transformed shapes is given by 

$$\langle \tilde{V}(u_j;\gamma) \rangle = \frac{1}{D_{max}-D_{min}}\sum\limits_{d=D_{min}}^{D_{max}} \tilde V_d(u_j;\gamma).$$

The quality of collapse is measured by a normalized mean squared error (NMSE) across the transformed durations:
\begin{equation}
\mathrm{NMSE}(\gamma)\;=\;\frac{\sum\limits_{j}\sum\limits_{D}\,\big[\tilde V_D(u_j;\gamma)-\langle \tilde{V}(u_j;\gamma) \rangle \big]^2}
{\sum\limits_{j}\,\big[\langle \tilde{V}(u_j;\gamma) \rangle\big]^2}.
\end{equation}

We perform a brute-force search to solve
\begin{equation}
\gamma^\star = \arg\min_{\gamma}\,\mathrm{NMSE}(\gamma).
\end{equation}

To estimate the uncertainty, we fit a quadratic function $a\gamma^2+b\gamma+c$ to $\mathrm{NMSE}(\gamma)$ in a small neighborhood of $\gamma^\star$ and use the curvature to compute
\begin{equation}
\sigma_\gamma = \sqrt{\frac{1}{2a}}.
\end{equation}

\section{Training and quasi-criticality on MNIST dataset}\label{ref: subsection MNIST methods}

Here we explore how initialization parameters $(\sigma_w^2,\sigma_b^2)$ affect the training performance of Gaussian initialized networks (See Text). We PyTorch library in Python and since deep learning computations can run in parallel across multiple CPU threads, we explicitly control this parallelism to ensure that every experiment was reproducible and independent of machine specific defaults. In practice, we set both the intra-op and inter-op thread counts to 16. The intra-op thread handles the number of threads used for parallelizing operations within a single operator, while the inter-op thread handles parallelism between different independent operators in the computation. Additionally, we align the numerical libraries OpenMP (OMP), Intel’s Math Kernel Library (MKL), NumExpr, and Apple’s VecLib, at the system level, by setting their maximum thread counts to 16 as well. This guarantees that performance differences across runs are attributable only to the neural network configuration, and not to varying levels of CPU parallelization. 

The MNIST handwritten digit dataset (Fig.~\ref{fig:susceptibility}) is divided in a training set containing 60000 grayscale images and a test set containing 10000 grayscale images, with a total of 70000 images. Each image has dimensions 28 $\times$ 28 pixels, with one unique grayscale channel, and there are 10 output possible classes, corresponding to the 10 digits, from 0 to 9 \cite{lecun1998mnist}. From the 60000 available training samples, we use a subset of 25600 images (200 batches × 128 images per batch) to reduce training time, while keeping results statistically meaningful. Computationally, each image is converted into a tensor, and then reshaped into a one-dimensional vector of length $28 \times 28 = 784$. 

The neural network architecture is defined by the following hyperparameters: input dimension, hidden dimension $(N)$, output dimension, and depth $(L+1)$ where the $L+1$-th layer is the output layer. The input layer, also noted as the 0-th layer, has 784 neurons, one for each pixel, connected to $N=300$ hidden neurons in the next layer. The output layer has 10 neurons, corresponding to the 10 possible digit classes. All hidden layers use the hyperbolic tangent ($\phi(.) = \tanh(.)$) activation function. In the experiments reported in this paper, dropout is not used.

We train the networks using stochastic gradient descent with a learning rate of $10^{-3}$, and a batch size of $128$, with cross-entropy loss function. Each experiment was run for $10$ epochs, where an epoch consisted of iterating over $200$ mini-batches from the chosen dataset subset. We quantified the performance as the training accuracy, that is, the fraction of correctly classified digits within those mini-batches. As our goal is not to benchmark MNIST but to directly measure how initialization parameters influence learning on the training data itself, we do not evaluate the performance on a validation or test set. In Fig.~\ref{fig:susceptibility}~(c) we show the accuracy reached for each $(\sigma_w^2, \sigma_b^2)$ initialization pair. 

Next, we extend our analysis by measuring how quickly a target level of performance is reached by the network. For that, we trained fully connected deep neural networks with a depth of $300$ layers and with varying hidden layer widths. In particular, we train networks with $200, 300, 400, 500, 600$ neurons per layer, initializing weights and biases as in Fig.~\ref{fig:susceptibility}~(c). In contrast to the previous experiments, here we fix the bias variance to $\sigma_b^2=0$ and we vary the weight variance $\sigma_w$ within a narrow range around the critical regime. For each configuration, we train the model for at most $50$ epochs and applied early stopping in case the network achieved a training accuracy of 97\% with less epochs. We report the number of epochs needed to reach 97\% accuracy in each case, up to maximum of 50 epochs. We train some of the networks that were not trainable within $50$ epochs for up to 200 epochs to ensure that learning was not happening even if the amount of epochs was significantly increased. Their accuracy remain low even with larger amount of epochs. In Fig.~\ref{fig:susceptibility}~(f) we report the number of epochs needed by different networks with different hidden dimensions, to achieve 97\% of training accuracy, for different values of $\sigma_w$ near the critical point, and fixing $\sigma_b^2=0$. The results show that trainability and, thus, learnability starts to be achieved for those networks initialized near the critical value $\sigma_w^2 \approx 1$, highlighting the dependency on the network's performance on the initialization. 

\begin{figure*}
    \centering
    \includegraphics[width=\linewidth]{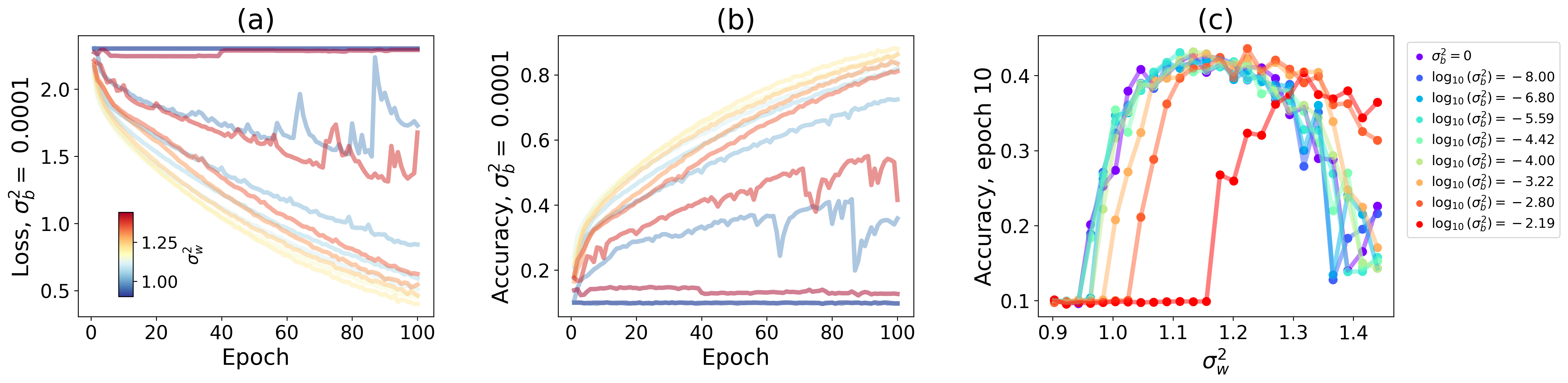}
    \caption{\arsham{\textbf{Training dynamics and initialization-dependent trainability on CIFAR-10.} (a) Training loss and (b) training accuracy across $100$ epochs for networks initialized with fixed bias variance $\sigma_b^2=10^{-4}$ and different weight variances $\sigma_w^2$, indicated by color. Only a restricted range of $\sigma_w^2$ produces sustained optimization and increasing accuracy, whereas initialization outside this region leads to poor trainability. (c) Training accuracy after $10$ epochs as a function of $\sigma_w^2$ for different bias variances $\sigma_b^2$. Increasing $\sigma_b^2$ shifts and progressively narrows the range of trainable weight variances.}}
    \label{fig:CIFAR}
\end{figure*}

\arsham{\section{Training and quasi-criticality on CIFAR-10 dataset}

We also examine the CIFAR-10 as a more challenging dataset, which contains $60000$ RGB images distributed across $10$ object categories, with $50000$ images used for training and $10000$ reserved for testing. Each image has resolution $32\times32$ with three color channels, corresponding to an input dimensionality of $32\times32\times3=3072$ after flattening. To make the experiments computationally tractable, we use a fixed subset of $38400$ training images, corresponding to $300$ mini-batches of size $128$. Images are converted to tensors, normalized using the standard CIFAR-10 channel statistics, and flattened into one-dimensional input vectors.

Similar to the MNIST setup, we consider a deep fully connected feedforward neural network with $N$ neurons per layer, depth $L+1$, input dimension, and output dimension. The input layer contains $3072$ units and is connected to hidden layers of width $N=300$. The network depth is fixed to $L+1=300$ (like in the MNIST case), resulting in $300$ successive linear transformations before the final classifier layer. The output layer contains $10$ units, one for each CIFAR-10 category. Hidden layers use $\tanh(x)$, without dropout regularization.

Optimization is performed using stochastic gradient descent with learning rate $10^{-3}$ and cross-entropy loss. Networks are trained for $100$ epochs with mini-batches of size $128$. Again, our objective is to investigate how the initialization parameters affect trainability in very deep networks and we report the training accuracy. While this dataset takes longer to train (see Fig.~\ref{fig:CIFAR}a and \ref{fig:CIFAR}b), it produces similar patterns to MNIST (Fig.~\ref{fig:susceptibility}). To better represent the similarity, we show the accuracy curves for all pairs of $(\sigma_w^2,\sigma_b^2)$ at epoch  10 (see Fig.~\ref{fig:CIFAR}c).
}

\section{ResNet statistics.}
Here, we elaborate on how we adapt our framework when working with non-Gaussian networks, specifically ResNet.

The heterogeneity of layer widths and operations in ResNets demands modifications in our method. Let $y^{\ell_b}\in\mathbb{R}^{H^{\ell_b}\times W^{\ell_b}\times C^{\ell_b}}$ denote the activation tensor received by block $\ell_b$, and let $z^{\ell_b}$ denote the pre-activation produced by the residual branch in block $\ell_b$. Note that we use $\ell_b$ to index block depth, while we enumerate the operations separately and by $\ell$ --- decomposing each block into multiple layers. Also, the number of channels is initially $C^{0}=3$ encoding the red, green, blue decomposition of images. Then, convolution modules mix channels in different ways, changing their number. 

ResNet dynamics can be compactly written as
\begin{eqnarray}
z^{\ell_b} &=& R^{\ell_b}\left(y^{\ell_b},\theta^{\ell_b}\right),
\label{eq:resnet-preact}\\
y^{\ell_b+1} &=& s^{\ell_b}\,y^{\ell_b} + \phi\big(z^{\ell_b}\big),
\label{eq:resnet-update}
\end{eqnarray}
where $s^{\ell_b}$ is the skip operator, letting a part of the previous block activations directly pass. $R^{\ell_b}(\cdot;\theta^{\ell_b})$ is the residual \emph{pre-activation} map with parameters $\theta^{\ell_b}$ (convolutional kernels, and BatchNorm scales and shifts) that for a \emph{basic} pre-activation block reads
\begin{eqnarray}
R^{\ell_b}(y) &=& \mathrm{BN}_{2}^{\ell_b}\left(\mathrm{Conv}_{2}^{\ell_b}\Big(\phi\big(\mathrm{BN}_{1}^{\ell_b}(\mathrm{Conv}_{1}^{\ell_b}(y^{\ell_b}))\big)\Big)\right),
\label{eq:R-basic}
\end{eqnarray}
where $\mathrm{Conv}_1$ and $\mathrm{Conv}_2$ are convolution layers, $\phi$ applies the Relu activation function and $\mathrm{BN}_1$ and $\mathrm{BN}_2$ are BatchNorms. Here we do not provide the full details of each ResNet structure and dynamics. However, we treat each operation, including maxpool, convolutions, non-linearities, batch normalization, as separate layers that produce tensors with $N^\ell$ components, whose norm gives the signal strength $\|x^\ell \| = \sqrt{\tilde{q}^\ell}$, where we use tilde to distinguish it from the analogous expression in the previous sections. This leads to the total maximum depths of $60, 158, 464 $ for ResNet models versions $18, 50, 152$, respectively. 

The heterogeneity of layer operations leads to highly variable signal strength--- especially the batchnorm operations that directly tune it. One way to overcome the complication is assigning layer dependent thresholds $\theta_{\ell} = \mu_{\ell} + n \sigma_{\ell} $, where $n$ is an integer and $\mu$ and $\sigma$ are mean and standard deviation of $\sqrt{q_{\ell}}$. We perform $10000$ runs to calibrate the threshold, prior to avalanche analysis that includes $1$ million perturbations for each model. We quantify the avalanche size as $\tilde{S} = \sum\limits_{\ell = 1}^{\ell_f-1} \sqrt{\tilde{q}_{\ell}} - \theta_{\ell}$, with $\ell_f$ being the layer at which signal strength goes below the threshold.

\section{Depth and performance}
\begin{figure*}[h!]
    \centering
    \includegraphics[width=\linewidth]{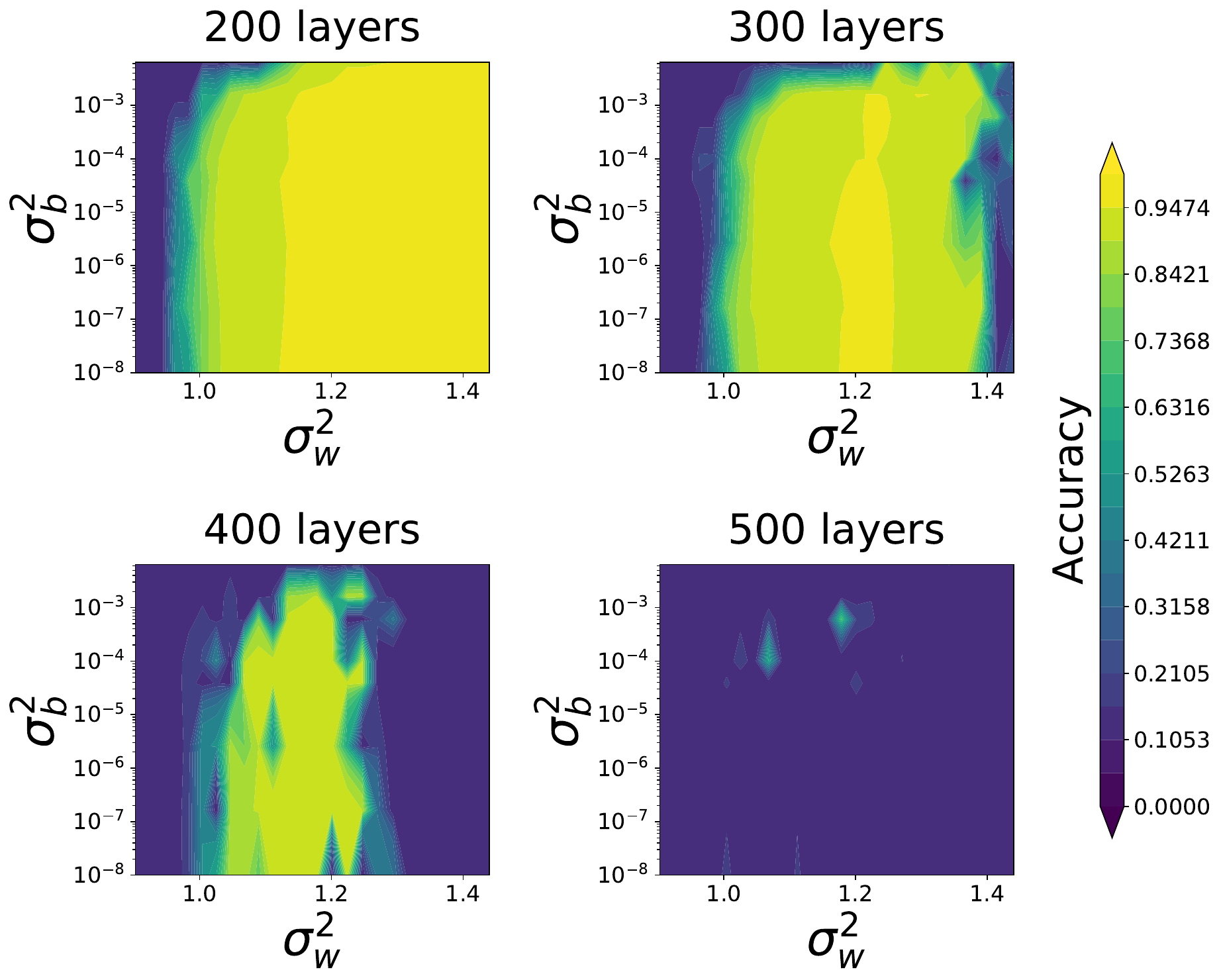}
    \caption{\textbf{Training accuracy for networks of increasing depth.} 
    Contour plots show the training accuracy obtained after 10 epochs on the MNIST dataset for fully connected deep neural networks with hidden dimension $N = 300$ and depths of 200, 300, 400, and 500 layers. Each point corresponds to the average accuracy over 200 mini-batches (128 samples each) for a given initialization pair $(\sigma_w, \sigma_b)$. The color scale represents the achieved accuracy. The squared bias variance $\sigma_b^2$ is shown on a logarithmic scale and $\sigma_w^2$ on a linear scale.}
    \label{fig:SI_depth200_to_500_sb_vs_sw}
\end{figure*}

Fig.~\ref{fig:SI_depth200_to_500_sb_vs_sw} extends the analysis presented in the main text by examining how network depth influences the relationship between initialization parameters and training performance. Using the same experimental setup (fully connected deep neural networks trained on a subset of the MNIST dataset with $N = 300$ hidden neurons per layer, 10 epochs of stochastic gradient descent, and no dropout), we varied the number of layers between 200, 300, 400, and 500. As depth increases, the regions of high training accuracy become progressively narrower, consistent with the increased sensitivity of very deep networks to initialization.

\end{document}